\documentclass[10pt]{iopart}
\usepackage{epsfig}
\begin{document}

\title[Stick boundary conditions and rotational velocity auto-correlation functions]{Stick boundary conditions and rotational  velocity auto-correlation
functions for colloidal particles in a coarse-grained representation
of the solvent} \author{J. T. Padding{\dag}, A. Wysocki{\ddag},
H. L{\"o}wen{\ddag} and A. A. Louis\dag} \address{ \dag Department of
Chemistry, University of Cambridge, Lensfield Road, Cambridge CB2 1EW,
United Kingdom} \address{\ddag Institut f{\"u}r Theoretische Physik
II, Heinrich-Heine-Universit\"at D{\"u}sseldorf,
Universit\"atsstra{\ss}e 1, D-40225 D\"usseldorf, Germany}
\begin{abstract}
We show how to implement stick boundary conditions for a spherical
colloid in a solvent that is coarse-grained by the method of
stochastic rotation dynamics. This allows us to measure colloidal
rotational velocity auto-correlation functions by direct computer
simulation.  We find quantitative agreement with Enskog theory for
short times and with hydrodynamic mode-coupling theory for longer
times.  For aqueous colloidal suspensions, the Enskog contribution to
the rotational friction is larger than the hydrodynamic one when the
colloidal radius drops below $35$nm.
\end{abstract}


\section{Introduction}

Colloidal particles exhibit Brownian motion due to collisions with the
molecules of the solvent in which they are suspended~\cite{Dhon96}.
The resulting momentum transfer leads to translational diffusion of
the colloidal positions. Because  real colloids are not perfectly
smooth or spherical, the solvent molecules can also transfer angular
momentum, leading to rotational diffusion around the colloidal
centres.  Computer simulations of this process are difficult for two
reasons: {\bf 1)} the length-scales of a typical colloid and a
solvent differ by many orders of magnitude, for example, a colloid of
radius $1 \mu$m displaces $1.4 \times 10^{11}$ water molecules,
making direct simulations  over meaningful length and timescales
impossible in practice. {\bf 2)} Colloid-solvent interactions are
usually taken to be radial, at least for a typical spherical colloidal
particle, which means that angular momentum is not transferred from
the solvent to the particle.  In other words, the traditional
implementation of molecular dynamics forces results in {\em slip
boundary conditions}~\cite{Schm03}, whereas for a realistic colloid,
local surface inhomogeneities would result in a tangential fluid
velocity at the surface equal to the local velocity of the colloid
surface, leading to {\em stick boundary conditions}~\cite{Bocq94}.

The solution to problem {\bf 1)} is to coarse-grain the fluid to
larger time and length-scales.  There are many different ways of doing
this, but in this paper we focus on stochastic rotation dynamics
(SRD), a method first introduced by Malevanets and Kapral in
1999~\cite{Male99}. SRD has the advantage that it includes thermal
effects that dominate the short-time velocity auto-correlation
functions, and also the hydrodynamic forces that dominate at longer
times.  Briefly, SRD represents the solvent by ideal particles.  To
coarse-grain collisions, space is partitioned into cubic cells and in
each one the particles exchange momentum with each other by rotating
their velocity around the centre of mass velocity of the cell during a
``collision'' step.  This procedure conserves momentum and kinetic
energy, and thus generates the correct thermal Navier Stokes solutions
in the thermodynamic limit.  Its simplicity means that transport
coefficients such as the viscosity can be calculated
analytically~\cite{Kiku03,Ihle04}, facilitating the choice of
simulation parameters.


Solute particles can also be coupled to the SRD solvent by treating
solvent-solute and solute-solute interactions by a standard molecular
dynamics scheme~\cite{Male00}.  This method leads to slip boundary
conditions for spherical colloids~\cite{Padd04,Lee04}, described as
problem {\bf 2)} above.  To study rotational correlations, stick
boundaries must be implemented, and how to do that is the subject of
the present paper.

We proceed as follows: In the section 2, we briefly review our
parameter choice for  SRD.  Section 3 describes how we implement stick
boundary conditions for colloids while section 4 contains our
simulation results for the linear and rotational velocity
auto-correlation functions that verify our stick boundary
implementation.

\section{SRD model parameters}

To simulate the SRD solvent, we follow our earlier implementation
described ref~\cite{Padd04}. Throughout this paper our results are
described in units of SRD mass $m$, SRD cell size $a_0$ and thermal
energy $k_B T$.  The number density (average number of SRD particles
per SRD cell) is fixed at $\gamma = 5$, the rotation angle is $\alpha
= \pi/2$, and the collision interval $\delta t_c = 0.1 t_0$, with 
time units $t_0 = a_0 (m/k_BT)^{1/2}$; this corresponds to a mean-free
path of $\lambda_{free} \approx 0.1 a_0$.  In our units these choices
mean that the the fluid viscosity takes the value $\eta=2.5 m/a_0t_0$
and the kinetmatic viscosity is $\nu=0.5 a_0^2/t_0$. The Schmidt
number Sc, which measures the rate of momentum (vorticity) diffusion
relative to the rate of mass transfer is given by Sc$ = \nu/D_f
\approx 5$, where $D_f$ is the fluid self-diffusion
constant\cite{Kiku03,Ihle04}.  In a gas Sc $ \approx 1$ -- momentum is
mainly transported by moving particles -- whereas in a liquid it is
much larger -- momentum is primarily transported by interparticle
collisions.  For our purposes it is only important that vorticity
diffuses faster than the particles do.  With these parameters the
Brownian time-scales are also well-separated\cite{Padd04}.  If $t_c$
is the time-scale over which the fluid loses memory of its velocity,
and $\tau_B$ is the time-scale over which the colloid loses memory of
its initial velocity, and $\tau_D$ is the time-scale over which the
colloid diffuses over one radius $a$, then for the smallest colloids
used, with radius $a=2$, $\tau_B \approx 20t_c$ and $t_B < \tau_D
\approx 200 t_c$.  For larger colloids the time-scales are separated
even further.  In contrast to earlier
papers~\cite{Male00,Padd04,Lee04}, the colloid-solvent interaction was
not treated by a smooth potential which leads to slip boundary
conditions.  Instead, an effective hard-sphere radius was imposed, and
the method of coupling the colloid to the solvent is described in the
next section.

\section{Stick boundary conditions for stochastic rotation dynamics}

Stick boundary conditions imply that the tangential fluid velocity
relative to an interface is zero at that interface~\cite{Bren83}.  The
detailed molecular origins of these boundary conditions, or even the
exact location at which they should be applied, are subtle
problems~\cite{Bocq94}. For a recent review of the extensive
literature on this subject see ~\cite{Laug05}. In some cases, such as
a non-wetting surface, large effective slip lengths can
occur~\cite{Barr99}, but for most colloidal applications the
length-scales over which these more complex processes occur are
coarse-grained out by methods such as SRD, so that simple stick
boundary conditions should be sufficient.

Bounce-back collision rules, where the tangential component of the
velocity relative to the surface of collision is reversed, could be
used to implement stick boundary conditions.  However, as Lamura {\em
et al}~\cite{Lamu01} showed in their study of Poiseuille flow, the
stochastic coarse-graining of the interparticle collisions together
with the grid shift necessary for Galilean invariance\cite{Ihle04} leads
to a finite slip length at a planar wall, even when bounce-back rules
are used. To fix this, they used the following prescription for cells
that intersect the boundary:
If number of SRD particles in a cell $n_{\mathrm{cell}}$ is smaller
than the average number of SRD particles per cell in the bulk
($\gamma$), they add the difference in virtual particles.  In practice
this means adding a Maxwellian velocity of variance $(\gamma -
n_{\mathrm{cell}})k_BT/m$ for each Cartesian component, and setting
the number density (for the calculation of the c.o.m.\ velocity) equal
to $\gamma$. If the number of SRD particles is larger than $\gamma$,
no virtual particles are added. We call this the Lamura rotation rule.
It increases the effective friction for a planar wall, and indeed
reduces the tangential velocity to nearly zero as required.  That
non-equilibrium effects can result in finite slip-lengths was shown as
early as 1879 by James Clerk Maxwell~\cite{Maxw79}, but his derivation
results in a slip-length proportional to the Knudsen number, and these
effects are an order of magnitude too small to explain the observed
slip.

An alternative to the bounce-back rules is to use stochastic
reflections where upon collision the particles are given a random
normal velocity $v_n$ and tangential velocity $v_t$ taken from the
following distributions:
\begin{eqnarray}
\label{eq_stochastic1}
P(v_n) & \propto & v_n \exp \left( -\beta v_n^2 \right) \\
P(v_t) & \propto & \exp \left( -\beta v_t^2 \right), \label{eq_stochastic2}
\end{eqnarray}
so that the wall acts as a thermostat~\cite{Lebo78}.  Real colloids
don't have  perfectly smooth surfaces: there could be a
grafted polymer brush for steric stabilisation, or an accumulation of
co- and counter-ions for charge-stabilised suspensions. Fluid
particles interacting with the surface would typically have multiple
collisions with these local inhomogeneities, and stochastic boundary
conditions can therefore be viewed as a coarse-grained representation
of these processes.  In that light, they appear more realistic than
bounce-back rules, but whether that is important for either long-ranged
hydrodynamic effects or for local Brownian motion is not clear.  Inoue
{\em et al}~\cite{Inou02} first implemented such stochastic boundary
conditions for SRD and more recently Hecht {\em et al}~\cite{Hech05}
used a similar, but more sophisticated, approach for spherical
colloids.

We applied the Lamura rotation rule along the curved surface of a
colloid (both reflection rules are equivalent), this is non-trivial
because it is not obvious how to choose the distribution of the
virtual particle velocities for a cell partially overlapping a
colloid.  All the implementations we tried resulted in rotational
frictions that were too large.  At present, the reasons for this are
not completely clear, they may be related to the fact that the
colloids can move, and therefore have a local temperature, in contrast
to the walls used in \cite{Lamu01}, which are immobile.

For spherical colloids we finally used the following implementation of
stochastic reflections, related to that used in \cite{Hech05}: If an
SRD particle overlaps with a colloid, go half a time step back
($-\mathbf{v}\,\, dt/2$), and place the particle along the shortest
vector $\mathbf{r}^*$ to the surface of the colloid. Then choose a
random velocity $\mathbf{v}'$ according to the stochastic
distributions Eqs.~(\ref{eq_stochastic1}) -- (\ref{eq_stochastic2}),
which are now centred around the local velocity of the colloid
surface, which is given by $\mathbf{v}_{\mathrm{loc}} = \mathbf{V} +
\mathbf{\Omega} \times \left( \mathbf{r}^* - \mathbf{R}\right)$, where
$\mathbf{\Omega}$ and $\mathbf{V}$ are the colloid rotation and
velocity vectors. Complete the second half of the time step with that
velocity ($+\mathbf{v}'\,\, dt/2$). Add all momenta changes for the
colloid: $\Delta \mathbf{P} = \sum m (\mathbf{v} - \mathbf{v}')$, as
well as all its angular momenta changes: $\Delta \mathbf{L} = \sum m
(\mathbf{r}^* -\mathbf{R}) \times (\mathbf{v}-\mathbf{v'})$. At the
end of the time step, update the linear and angular velocity of the
colloid according to $\mathbf{V} \rightarrow \mathbf{V}+\Delta
\mathbf{P}/M$ and $\mathbf{\Omega} \rightarrow \mathbf{\Omega}+\Delta
\mathbf{L}/I$, where $M$ is the mass and $I = 2/5\, Ma^2$ the moment of
inertia a colloid for radius $a$.  This method takes into account the
fact that, on average, crossings take place at half a time-step. It
proved to work slightly better than resetting the particle to the
surface immediately, and then moving it with its new velocity for a
full time step as done in ref.~\cite{Hech05}.  An exact calculation of
impact time and place would be even better but this is computationally
much slower. In any case, for small integration steps, there is not
much difference between these different methods.

Finally, bounce back reflections can also be implemented, and give
similar results to stochastic boundary conditions, but the latter were
preferred because they locally thermostat the fluid, which may be
important under flow conditions.

\begin{figure}[ht]
\scalebox{0.6}{\includegraphics{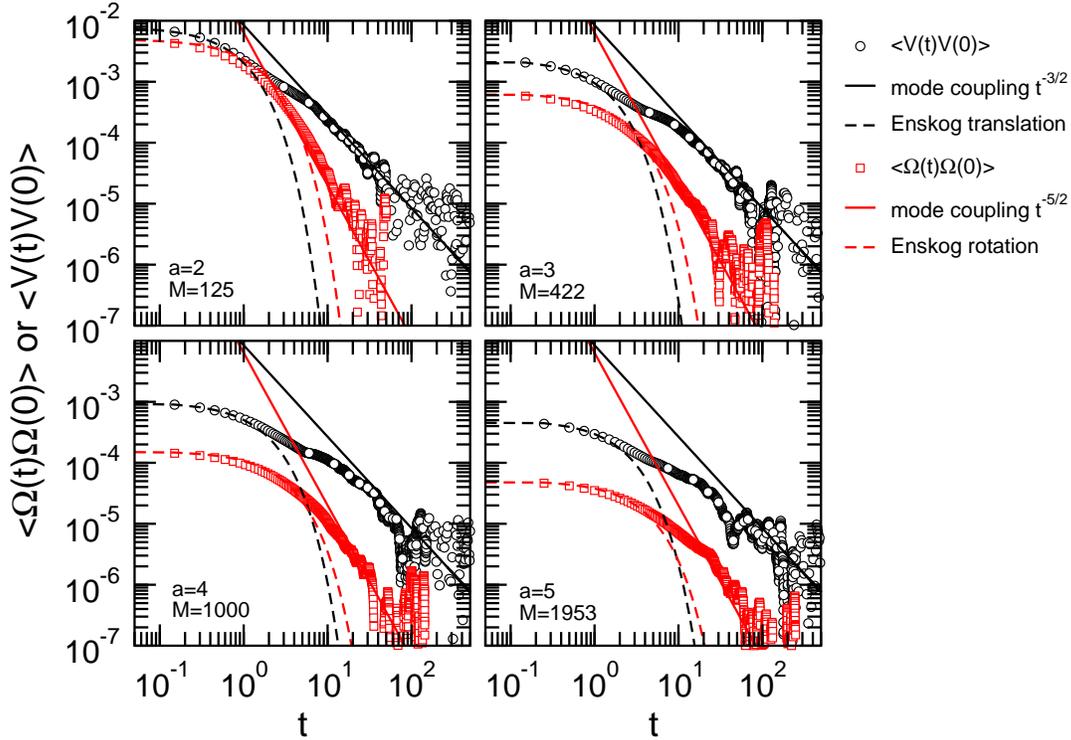}}
\label{fig_autocorrelations}
\caption{(colour online) Translational (circles) and rotational
 (squares) velocity autocorrelations for four different colloid radii.
 The x-axes are in units of $t_0= a_0 (m/k_BT)^{1/2}$, and the
 y-axes are in absolute units, e.g.$\left<V(0)V(0)\right> = k_BT/M =
 1/125$ for $a=2$ etc\ldots.  Solid lines are hydrodynamic
 mode-coupling predictions from
 Eqs.~(\protect\ref{eq_hydro1})--(\protect\ref{eq_hydro2}), dashed
 lines are short time Enskog friction predictions from
 Eqs.~(\protect\ref{eq_Enskog})--(\protect\ref{eq_Enskog2}), and show
 good agreement with no free parameters.  Note that for the smaller
 colloids, the Enskog and hydrodynamic regimes overlap substantially.}
\end{figure}

\section{Translational and rotational correlation functions}

As a test of our stick-boundary conditions, we directly measure the
translational velocity autocorrelation function (VACF), defined as
$\left\langle V(t) V(0) \right\rangle$, where $V(t)$ is a Cartesian
component of the translational velocity of the colloidal particle at
time $t$, as well as the rotational velocity autocorrelation (RVACF),
defined as $\left\langle \Omega(t)\Omega(0) \right\rangle$, where
$\Omega(t)$ is a component of its rotational velocity.  At $t=0$,
equipartition yields $\left\langle V^2 \right\rangle = k_BT/M$ and
$\left\langle \Omega^2 \right\rangle = k_BT/I$.

For short times, the correlation functions can be approximated using
Enskog dense-gas kinetic theory, see e.g.~\cite{Subr75,Hyne77}, which 
predicts the following exponential decay:
\begin{eqnarray}\label{eq_Enskog}
\lim_{t \to 0} \left\langle V(t) V(0) \right\rangle &=& \left\langle V^2 \right\rangle
\exp \left( -\zeta_{\mathrm{ENS}}^V t \right) \\
\lim_{t \to 0} \left\langle \Omega(t) \Omega(0) \right\rangle &=&
\left\langle \Omega^2 \right\rangle
\exp \left( -\zeta_{\mathrm{ENS}}^{\Omega} t \right),
\end{eqnarray} 
where the Enskog friction coefficients are given by
\begin{eqnarray}
\zeta_{\mathrm{ENS}}^V & = & \frac{8}{3}
\left( \frac{2 \pi k_BT m M}{m + M} \right)^{1/2}
\frac{1}{M} \gamma a^2 \frac{1+2\chi}{1+\chi} \\
\zeta_{\mathrm{ENS}}^{\Omega} & = & \frac{8}{3}
\left( \frac{2 \pi k_BT m M}{m + M} \right)^{1/2}
\frac{1}{M} \gamma a^2 \frac{1}{1+\chi} \label{eq_Enskog2}
\end{eqnarray}
and $\chi = I/(Ma^2) = 2/5$ is the gyration ratio. Note that
$\zeta_{\mathrm{ENS}}^V / \zeta_{\mathrm{ENS}}^{\Omega} = 1 + 2\chi =
9/5 > 1$, so that the short time decorrelation of linear velocity is
{\em faster} than the short time decorrelation of angular velocity,
contrary to what would happen if one (erroneously) described the Brownian
relaxation with hydrodynamic friction coefficients:
\begin{eqnarray}
\zeta_h^V & = & 6 \pi \eta a / M \\
\zeta_h^{\Omega} & = & 8 \pi \eta a^3 / I. 
\end{eqnarray}
Using only these Stokes-Einstein frictions, one finds $\zeta_h^V /
\zeta_h^{\Omega} = 6 \chi /8 = 3/10 < 1$, which gives exactly the
opposite effect (a difference of a factor $6$)!

For the long time relaxation we can compare with hydrodynamic
mode-coupling theory~\cite{Erns70,Mast85} which predicts
algebraic long-time tails of the form:
\begin{eqnarray}\label{eq_hydro1}
\lim_{t \to \infty} \left\langle V(t) V(0) \right\rangle &=&
\frac{k_BT}{12 m \gamma \left( \pi (\nu + D_c) t \right)^{3/2}} 
\\
\lim_{t \to \infty} \left\langle \Omega(t) \Omega(0) \right\rangle &=&
\frac{k_BT \pi}{m \gamma \left(4 \pi (\nu + D_c) t\right)^{5/2}}
\label{eq_hydro2}
\end{eqnarray}
where $D_c$ is the colloidal diffusion constant which, in this case,
is much smaller than the kinematic viscosity $\nu$, and can be
ignored.

In Figure \ref{fig_autocorrelations} we show VACF's and RVACF's for
colloids of various radius $a$.  We kept the colloid density constant
(i.e. $M \propto a^3$) and investigated four different sizes: $a=2$
$(M=125)$, $a=3$ $(M=422)$, $a=4$ $(M=1000)$, and $a=5$ $(M=1953)$.
To keep finite-size corrections~\cite{Loba04} small and comparable in
each case, the box-size was varied to keep $L/a=24$ fixed.  Thus the
number of SRD particles varied from about $5.5 \times 10^5$ for $a=2$
to $8.6 \times 10^6$ for $a=5$.  We find excellent agreement with
theory for the short-time behaviour, and good agreement for the
long-time behaviour of the autocorrelation functions.  For the smaller
particles the statistics for the long-time tails are better than for
the larger particles.  Overall, these results suggest that our
coarse-grained method to implement stick boundary conditions leads to
the correct physical behaviour.

Note that for the smallest colloid, the Enskog and hydrodynamic
friction regimes are not clearly separated, especially for the
rotational autocorrelation function.  A naive parallel
addition of the two types of friction to find the total friction
$\zeta$:
\begin{equation}\label{eq_friction}
\frac{1}{\zeta} = \frac{1}{\zeta_{ENS}} + \frac{1}{\zeta_h}
\end{equation}
will therefore be incorrect, in particular for the rotational friction
coefficient.  For larger particles, this works better though, as shown
in~\cite{Lee04} where a similar analysis is performed for the
translational friction coefficient of an colloidal particle with slip
boundary conditions.  Naively using only the Stokes-Einstein
hydrodynamic frictions, i.e.\ assuming that the hydrodynamic radius is
equal to the hard-sphere radius, will lead to important errors for
small colloids. Consider the following example of a density matched
colloid in an H$_2$0 solvent with viscosity $\eta = 0.001$ Pa s at a
temperature of $300$K.  Taking $a$ in meters and $\zeta$ in s$^{-1}$,
one finds $\zeta^V_{ENS} = 764/a$ and $\zeta^V_h = 4.5\times
10^{-6}/a^2$ for translations and $ \zeta^\Omega_{ENS} = 5/9
\,\,\zeta^V_{ENS} = 424/a$ and $\zeta^\Omega_h = 10/3 \,\, \zeta^V_h =
15\times 10^{-6}/a^2$ for rotations.  The physical radius at which the
Enskog and hydrodynamic frictions are equal is given then by
$a_{crit}^V = 6$nm for translations and $a_{crit}^\Omega=35.4$ nm for
rotations.  Because $\zeta_{ENS}/\zeta_{h} \sim a$, even for $100$ nm
the Enskog contribution to the rotational friction is of order $30\%$
and cannot be ignored\footnote{The quantitative agreement with Enskog
theory we find here may be due to the weaker correlations in the SRD
fluid as compared to a full microscopic fluid.  Because the SRD fluid
and the boundary conditions are idealised, the exact radius where the
microscopic Enskog friction and the hydrodynamic friction are equal
will depend on further details of realistic physical systems.}.

\section*{Conclusions}

We have shown how to implement stick boundary conditions for a colloid
in a coarse-grained SRD solvent. Stochastic reflections with an
approximate rule to determine the point where the SRD particle crossed
the colloid surface was found to work best.  This method was tested by
explicit computer simulations of the translational and rotational
velocity autocorrelation functions, which compare well to analytic
calculations of the short-time behaviour via Enskog theory, and the
long-time behaviour via mode-coupling theory.  Our successful
implementation of stick boundary conditions also shows that for small
particles the Enskog and hydrodynamic effects are not clearly
separated, and that the Enskog, or microscopic
contribution\cite{Hyne77}, can be larger than the hydrodynamic one for
small (nano) colloids, in particular for the rotational friction.

\section*{Acknowledgements}
JTP acknowledges support from the EPSRC and IMPACT FARADAY, and
Schlumberger Cambridge Research. AAL acknowledges support from the
Royal Society (London). AW and HL thank the the Deutsche
Forschungsgemeinschaft (DFG), for support through the SFB-TR6 program
`Physics of colloidal dispersions in external fields'.

\end{document}